\documentclass[aps,prl,twocolumn,superscriptaddress,floatfix]{revtex4}
\usepackage{graphicx}
\usepackage{algorithmicx,algorithm,algpseudocode}
\usepackage{amsmath,mathtools,amsfonts}
\usepackage{hyperref}
\usepackage{microtype}

\newcommand{\tr}{\mbox{\textrm{Tr}}}

\newcommand{\york}{Department of Physics and Astronomy, York University, 
Toronto, Ontario, M3J 1P3, Canada}

\begin{document}

\title{Fourier Accelerated Conjugate Gradient Lattice Gauge Fixing}

\author{R.J. Hudspith}
\affiliation{\york}

\collaboration{The RBC and UKQCD Collaborations}\noaffiliation

\begin{abstract}  
We provide details of the first implementation of a non-linear conjugate gradient method for Landau and Coulomb gauge fixing with Fourier acceleration. We find clear improvement over the Fourier accelerated steepest descent method, with the average time taken for the algorithm to converge to a fixed, high accuracy, being reduced by a factor of 2 to 4.
\end{abstract}

\maketitle

\section{Introduction}\label{sec:intro}

Conjugate gradient (CG) methods (first used to solve linear equations \cite{zbMATH03075195} and later generalised for non-linear, non-quadratic functions \cite{Fletcher01011964,Polyak196994}) are a technique to solve unconstrained local minimisation problems. Numerically these methods can be implemented cheaply because they are iterative and converge in a finite number of steps. They are also often considered computationally faster than the steepest descent method \cite{Press:1992:NRC:148286}. We illustrate how fixing to the smooth, Landau and Coulomb gauges in the context of lattice field theory can be achieved by using the method of conjugate gradients.

Fixing the gauge is a prescription for removing redundant degrees of freedom of the gauge field in a continuum quantum field theory. Common choices are the Landau $\partial_\mu A_\mu(x)=0$ and Coulomb $\partial_i A_i(x)=0$ gauges (where Greek indices run over all dimensions and Roman over the spatial and the $A$'s are the gauge fields of our theory). Fixing the gauge (while not necessary for many lattice measurements) is often required for the direct matching of lattice simulations to continuum perturbation theory.

Measurements of lattice Green's functions in strongly coupled, $N_d$-dimensional, $\text{SU}(N_c)$ theories have to be performed with a fixed gauge and are often computed in Landau gauge. Landau gauge Green's functions are vital for the non-perturbative renormalisation of important physical quantities such as the QCD Kaon bag parameter $B_K$ \cite{Aoki:2010pe,Boyle:2012qb} and can also be used for the measurement of the QCD strong coupling $\alpha_s$ \cite{Blossier:2013ioa}. Coulomb gauge fixing is more generally applicable to lattice theorists; it is used in methods such as gauge-fixed wall source quark correlators (also often used in the calculation of $B_K$), or for computing the static quark potential \cite{Kaczmarek:2002mc}. Having a fast routine to fix the gauge allows for faster measurements of critical physical quantities.

Lattice ``links'' transcribe the gauge fields via the matrix exponential (for lattice site ``$x$'', with lattice spacing $a$ and bare coupling $g_0$),
\begin{equation}
 U_\mu\left(x+a\frac{\hat\mu}{2}\right) = e^{iag_0 A_\mu\left(x+a\frac{\hat\mu}{2}\right)}.
\end{equation}

The gauge fields are obtained by the logarithm of the links. A common approximation \cite{Mandula1987127} to the logarithm of the map $U=e^{iA},U\in \text{SU}(N_c)$ is what we call the ``Hermitian projection'' (where $I_{N_c\times N_c}$ is the identity matrix),
\begin{equation}\label{eq:hermproj}
A = \frac{1}{2i}\left\{\left[U-U^{\dagger}\right]-\frac{1}{N_c}\tr\left[ U-U^{\dagger}\right]\cdot I_{N_c\times N_c}\right\}.
\end{equation}
This definition is not unique, being correct up to terms of $O(A^3)$. Exact logarithm techniques are possible \cite{Durr:2007cy,Ilgenfritz:2010gu} but will not be discussed here as they are numerically costly to implement and less commonly used in practice.

\section{Gauge fixing on a Lattice}\label{sec:generalities}

We now discuss the case of lattice Landau gauge fixing, as the extension to Coulomb gauge should be simple. After we introduce the conjugate gradient procedure we then discuss our implementation for fixing to Coulomb gauge.

There are two common types of gauge fixing routines, the Los Alamos \cite{Mandula1987127} and the Cornell \cite{PhysRevD.37.1581}, both of which use the method of steepest descent to minimise the functional (where $V$ is the lattice volume),
\begin{equation}\label{eq:funcmin}
 F(U) = \frac{1}{N_d N_c V}\sum_{x,\mu} \tr\left[\left(ag_0 A_\mu\left( x+a\frac{\hat\mu}{2}\right)\right)^2\right].
\end{equation}
We focus on the Cornell method as it can be Fourier accelerated.

For the Hermitian projection definition (Eq.\ref{eq:hermproj}) of the gauge fields, the following approximation of the functional can be used,
\begin{equation}\label{eq:trace_functional}
 F(U) \approx 1 - \frac{1}{N_d N_c V}\sum_{x,\mu}\Re\left(\tr\left[ U_\mu\left(x+a\frac{\hat\mu}{2}\right)\right]\right).
\end{equation}

In general, the method of steepest descent is a technique to find a local minimum of a function. Considering the n$^{\text{th}}$ iteration of such a method, the update,
\begin{equation}\label{eq:steep}
x_{n+1} = x_{n} - \alpha f^{\prime}(x_{n}),
\end{equation}
will step toward a local minimum of the function $f(x)$ provided the parameter is small, $0<\alpha<1$.

In direct analogy to the general procedure of Eq.\ref{eq:steep} we first approximate the derivative of a gauge field by,
\begin{equation}
 a\Delta_\mu A_\mu(x) = \sum_\mu \left( A_\mu\left( x+a\frac{\hat\mu}{2}\right) - A_\mu\left( x-a\frac{\hat\mu}{2}\right) \right).
\end{equation}
The $n^{\text{th}}$ iteration of the lattice steepest descent Landau gauge fixing procedure updates the links via the gauge transformation,
\begin{equation}\label{eq:gtrans_landau}
\begin{gathered}
g(x) = e^{-i\alpha a\Delta_\mu ag_0 A^{(n)}_\mu(x)}, \\ 
U^{(n+1)}_\mu\left(x+a\frac{\hat\mu}{2}\right) = g(x) U^{(n)}_\mu\left(x+a\frac{\hat\mu}{2}\right)g(x+a\hat\mu)^{\dagger}.
\end{gathered}
\end{equation}
This procedure successively minimises $a\Delta_\mu ag_0A_\mu^{(n)}(x)$, whilst retaining the gauge invariance of the action. The parameter $\alpha$ is again a small tuning parameter, which could be tuned at each step but is best fixed to a near-optimal constant value. We found setting $\alpha$ to 0.08 and 0.1 for Landau and Coulomb gauge respectively to be best for the ensembles considered in this work.

The exponential in Eq.\ref{eq:gtrans_landau} could be computed exactly using the technique of \cite{PhysRevD.69.054501}, but expansion to the term linear in $\alpha$ and reunitarisation is sufficient and numerically faster.

It is common to stop the gauge fixing routine once the quantity,
\begin{equation}
 \Theta^{(n)} = \frac{1}{N_c V}\sum_x \tr\left[ \left( a\Delta_\mu ag_0 A_\mu^{(n)}(x)\right)^2 \right],
\end{equation}
reaches some small value (often $\Theta^{(n)} \approx 10^{-14}$).

\section*{Fourier acceleration}

The steepest descent method of fixing to Landau gauge as outlined above was shown in \cite{PhysRevD.37.1581} to suffer from critical slowing down, meaning that the number of iterations required to converge to a fixed accuracy grows drastically with the volume of the problem. Their method to ameliorate this was to apply a re-scaling in momentum space of the eigenvalues of the (Abelian) Laplacian $\Delta^2$,
\begin{equation}\label{eq:FAgtrans}
 g(x) = e^{-i\alpha \: \tilde{F} \: \frac{p^2_{\text{Max}}}{V p^2} \: F \: a\Delta_\mu ag_0 A_\mu^{(n)}(x)},
\end{equation}
where $F$ and $\tilde{F}$ are forward and backward fast Fourier transforms (FFTs) respectively, and the factor of $\frac{1}{V}$ is for the FFT normalisation. In practice, it is best if the quantities $\frac{p^2_{\text{Max}}}{Vp^2}$ are precomputed as a look up table. The discrete momenta $p^2$ have the usual lattice definition,
\begin{equation}\label{eq:psq}
 p^2 = 2\left( N_d - \sum_\mu \cos\left( \frac{2\pi n_\mu}{L_\mu}\right) \right),
\end{equation}
with Fourier modes $n_\mu = \left( \frac{-L_\mu}{2},...,-1, 0 ,1, ... ,\frac{L_\mu}{2}-1 \right)$. $L_\mu$ is the length of the lattice in the $\mu$ direction. Special care should be taken at the zero mode, where we set the value of $p^2$ to 1 \cite{PhysRevD.37.1581}. 

In our implementation of Fourier acceleration, the shared-memory parallel version of the library FFTW \cite{FFTW05} was used. The Fourier accelerated steepest descent method will be denoted as FASD later on in this paper.

\section{The conjugate gradient method}

An outline of the general approach for the non-linear (Polyak-Ribier\'e \cite{Polyak196994})\footnote{We find that the Polyak-Ribier\'e definition of $\beta_n$ reduces the iteration count in comparison to the Fletcher-Reeves.} conjugate gradient method is shown in Alg.\ref{alg:CG}.

\begin{algorithm}
  \caption{General non-linear CG}
  \label{alg:CG}
  \begin{algorithmic}
  \State $\text{Compute the gradient direction }f^{\prime}(x_0)$%
  \State Perform a line search for $\alpha_0$ s.t min$(f(x_0 -\alpha_0 f^{\prime}(x_0)))$
  \State Perform the update $x_1 = x_0 - \alpha_0 f^{\prime}(x_0)$%
  \State Set $s_0 = -f^{\prime}(x_0)$%
  \State $n = 1$
  \While { $|f^{\prime}(x_n)|^2 >\text{Tolerance} $}
       \State Compute the gradient $f^{\prime}(x_n)$
       \State Compute $\beta_n= \text{max}\left[0,\frac{f^{\prime}(x_n)^T \left( f^{\prime}(x_n) - f^{\prime}(x_{n-1})\right)}{f^{\prime}(x_{n-1})^T f^{\prime}(x_{n-1})} \right]$
       \State Compute conjugate direction $s_{n} = -f^{\prime}(x_n) + \beta_n s_{n-1}$
       \State Perform a line search for $\alpha_n$ s.t min$(f(x_{n}+\alpha_n s_n))$
       \State Update $x_{n+1} = x_{n}+ \alpha_n s_n$
       \State $n = n+1$
  \EndWhile
  \end{algorithmic}
\end{algorithm}

The translation of this approach to lattice Landau gauge fixing follows almost directly, and we call it the Fourier Accelerated Conjugate Gradient (FACG) method, and outline it in Alg.\ref{alg:CG_FALAN}. This algorithm should not be confused with the CGFA algorithm of \cite{Cucchieri:1998ew}, which uses the CG algorithm to invert the Laplacian instead of performing FFTs.

The approach begins with an FASD step as in Eq.\ref{eq:FAgtrans}, storing the result of the Fourier accelerated derivative in the conjugate direction $s_n(x)$. Once the algorithm has reached a sufficient minimum we are finished, otherwise we repeat the procedure generating conjugate directions weighted by the factor $\beta_n$.

We choose to use a line search to approximately determine the optimal tuning parameter $\alpha_n$ at each iteration. To do so we evaluate the gauge fixing functional (Eq.\ref{eq:trace_functional}) for possible fixed probe values of the parameter $\alpha$, which we call $\alpha^{\prime}$ and create a cubic spline interpolation of the result. We then solve for the exact minimum of the cubic spline.

The evaluation of each probe $\alpha^{\prime}$ is the most expensive aspect of this approach as they each require both the exponentiation of the derivatives into the gauge transformation matrices, and a subsequent gauge transformation over the whole lattice. Therefore, performing the bare minimum number of evaluations in this step is key to a fast implementation of this procedure. 

For both Landau and Coulomb gauge, the probes $\alpha^{\prime}\in(0.0,0.15,0.3)$ were used. The point at $\alpha^{\prime}=0$ is the cheapest probe to evaluate because no extra exponentiations and gauge transformations are required; it should always be used.

\begin{algorithm}
  \caption{Landau gauge FACG}
  \label{alg:CG_FALAN}
  \begin{algorithmic}
  \State $\Gamma_0(x) \gets \: \tilde{F} \: \frac{p^2_{\text{Max}}}{V p^2} \: F \: a\Delta_\mu ag_0 A_\mu^{(0)}(x)$ %
  \State $g(x) \gets e^{-i\alpha_0 \Gamma_0(x) }$%
  \State $U_{\mu}^{(1)}\left(x+a\frac{\hat\mu}{2}\right)\gets g(x)U_{\mu}^{(0)}\left(x+a\frac{\hat\mu}{2}\right)g(x+a\hat\mu)^{\dagger}$%
  \State $s_0(x) \gets \Gamma_0(x) $%
  \State $n = 1$
  \While { $\Theta^{(n)} >\text{Tolerance}$ }
      \State $\Gamma_n(x) \gets \tilde{F} \frac{p^2_{\text{Max}}}{Vp^2} F a\Delta_\mu ag_0 A^{(n)}_\mu(x)$%
      \State $\beta_n = \text{max}\left[0,\frac{\sum_x \tr\left[ \Gamma_n(x)\left( \Gamma_n(x) - \Gamma_{n-1}(x) \right) \right]}{\sum_x \tr\left[ \Gamma_{n-1}(x) ^2 \right]} \right]$%
      \State $s_n(x) \gets \Gamma_n(x) + \beta_n s_{n-1}(x)$%
      \State $\alpha_n = \text{min}( F(U) ; \alpha^{\prime} s_n )$%
      \State $g(x) \gets e^{-i\alpha_n s_n(x)}$%
      \State $U_{\mu}^{(n+1)}\left(x+a\frac{\hat\mu}{2}\right) \gets g(x) U_{\mu}^{(n)}\left(x+a\frac{\hat\mu}{2}\right) g(x+a\hat\mu)^{\dagger}$%
      \State $n = n+1$
  \EndWhile
  \end{algorithmic}
\end{algorithm}

The function $\text{min}( F(U) ; \alpha^{\prime} s_n )$ finds the value of $\alpha_n$ such that
\begin{equation}
\begin{aligned}
 1-\frac{1}{N_d N_cV}\sum_{x,\mu} &\Re\bigg( \tr\bigg[ e^{-i\alpha_n s_n(x)}\\
 &U_{\mu}^{(n)}\bigg(x+a\frac{\hat\mu}{2}\bigg) e^{i\alpha_n s_n(x+a\hat\mu)}\bigg]\bigg),\nonumber
 \end{aligned}
\end{equation}
is approximately best minimised. 

As an optimisation, due to the Hermiticity of the gauge fields and hence their derivatives, we only need to Fourier transform the upper or lower triangular part of the matrix $a\Delta_\mu ag_0 A_\mu(x)$. The rest of the matrix can be reconstructed by Hermiticity and the final element of the matrix by tracelessness. Not only does this reduce the number of FFTs the routine performs but it also reduces the memory required to store these matrices.

A negative aspect of this method is that extra storage of the matrices $\Gamma_{n}(x)$ and $s_n(x)$ has to be made available, although these can be stored in the shortened form as mentioned above.

As the gauge fixing progresses and $\Theta^{(n)}$ approaches zero, the functional (Eq.\ref{eq:trace_functional}) flattens so that within finite numerical precision the minimum is indistiguishable for any probe $\alpha^{\prime}$. Once we reach this accuracy (approximately the machine epsilon of the storage precision of our fields) we switch to the Fletcher-Reeves \cite{Fletcher01011964} definition of $\beta_n=\frac{\sum_x \tr\left[\Gamma_n^2\right]}{\sum_x \tr\left[\Gamma_{n-1}^2\right]}$ and fixed $\alpha_n$ (i.e. we turn off the line search altogether).

\section{Coulomb gauge}\label{sec:coulomb}

Fixing to Coulomb gauge on the lattice amounts to the minimisation of the functional,
\begin{equation}\label{eq:coul_func}
 F(U) = \frac{1}{(N_d-1)N_c V}\sum_{x,i} \text{Tr}\left[ \left( ag_0 A_i\left(x+a\frac{\hat i}{2} \right) \right)^2\right],\nonumber
\end{equation}
where the Roman index ``$i$'' runs over the spatial indices only. The functional is time-slice independent; by exploiting this a time-slice by time-slice scheme can be a very effective method in fixing to Coulomb gauge. 

It is observed that each time-slice takes a greatly different number of iterations to converge compared to one another \cite{Nakagawa:2009zf}. If we used a lattice-wide technique, we could spend a lot of time waiting for the slowest converging time-slices as well as requiring $N_d$-dimensional, volume-wide FFTs. This advocates an approach which treats each time-slice separately. We will detail our implementation of a time-slice by time-slice FASD procedure as extension of this to an FACG procedure is straightforward.

We define the lattice gauge field derivative for fields restricted to a time-slice ``$t$'',
\begin{equation}
 a\Delta_i A_i(x,t) = \sum_i \left(A_i\left( x + a\frac{\hat i}{2} ,t\right) - A_i\left( x - a\frac{\hat i}{2} ,t\right)\right).\nonumber
\end{equation}

The $n^{\text{th}}$ iteration of the FASD method for Coulomb gauge fixing requires the computation of the gauge transformation matrices ($g^{(0)}(x,t)=I_{N_c\times N_c}$),
\begin{equation}\label{eq:accum}
 g^{(n+1)}(x,t) = e^{-i\alpha\: \tilde{F} \: \frac{p^2_{\text{Max}}}{V_{N_d-1} p^2} \: F \: a\Delta_i ag_0 A_i^{(n)}(x,t) }g^{(n)}(x,t).
\end{equation}
The Fourier transforms and momenta are defined in the $N_d-1$ dimensional (spatial) subspace whose volume is denoted $V_{N_d-1}$, with $p^2_{\text{Max}}=4(N_d-1)$.

As expressed in Eq.\ref{eq:accum}, we store the gauge transformation matrices as an accumulated product of the previous iterations. Therefore, each computation of the derivatives requires a local gauge transformation of the original links, and we do not overwrite the link matrices on that time-slice until the routine has converged. The Hermitian projection definition of our fields (Eq.\ref{eq:hermproj}) then reads (where the subscript $_\text{Trf}$ refers to the trace subtraction),
\begin{equation}\label{eq:spatial_gtrans}
   \begin{aligned}
    ag_0 &A_i^{(n)}\left(x+a\frac{\hat{i}}{2},t\right) = \\
      \frac{1}{2i}\bigg[& g^{(n)}(x,t)U_i\left(x+a\frac{\hat{i}}{2},t\right)g^{(n)}(x+a\hat i,t)^{\dagger} - \\
      &g^{(n)}(x+a\hat i,t) U_i\left(x+a\frac{\hat{i}}{2},t\right)^{\dagger}g^{(n)}(x,t)^{\dagger} \bigg]_{\text{Trf}}.
  \end{aligned}
\end{equation}

We use the criterion (where the sum is over the fields on a time-slice),
\begin{equation}
\theta^{(n)}(t) = \frac{1}{N_c V_{N_d-1}} \sum_{x} \text{Tr}\left[ \left( a\Delta_i ag_0 A_i^{(n)}(x,t) \right)^2 \right],\nonumber
\end{equation}
to estimate our convergence.

Once we have converged the gauge transformation matrices for a time-slice $g(x,t)$ and the ones above it $g(x,t+a)$, we can overwrite the link matrices by gauge transformation,
\begin{equation}\label{eq:gtrans_method}
 \begin{aligned}
  U_i\left(x+a\frac{\hat{i}}{2},t\right) &\leftarrow g(x,t) U_i\left(x+a\frac{\hat{i}}{2},t\right) g(x+a\hat i,t)^{\dagger}, \\
  U_t\left(x,t\right) &\leftarrow g(x,t) U_t\left(x,t\right) g(x,t+a)^{\dagger}.
 \end{aligned}
\end{equation}
This ensures the gauge invariance of the action under this procedure.

The algorithm for our procedure is presented in Alg.\ref{Alg:CFASD}, illustrating the use of temporary, time-slice wide, gauge transformation matrices (called $p,q$ and $r$) as a computer memory saving practice. It should be noted that the spatial gauge transformations for the $A_i^{(n)}$'s (Eq.\ref{eq:spatial_gtrans}) are now performed in terms of the $p$'s, $q$'s and $r$'s within the respective ``while'' loops.

\begin{algorithm}
  \caption{Slice-by-slice Coulomb gauge FASD}
  \label{Alg:CFASD}
  \begin{algorithmic}
  \State $n = 0 , p(x) \gets I_{N_c\times N_c}$%
  \While {$\theta^{(n)}(0) > \text{Tolerance}$}
  \State $p(x) \gets  e^{-i\alpha\: \tilde{F} \: \frac{p^2_{\text{Max}}}{V_{N_d-1}p^2} \: F \: a\Delta_i ag_0 A_i^{(n)}(x,0) }p(x)$%
  \State $n = n+1$
  \EndWhile
  \State $n = 0, q(x) \gets I_{N_c\times N_c}$%
  \While {$\theta^{(n)}(1) > \text{Tolerance}$}
  \State $q(x) \gets  e^{-i\alpha \: \tilde{F} \: \frac{p^2_{\text{Max}}}{V_{N_d-1}p^2} \: F \: a\Delta_i ag_0 A_i^{(n)}(x,1) }q(x)$%
  \State $n = n+1$
  \EndWhile
  \State $U_{i}\left(x+a\frac{\hat{i}}{2},0\right) \gets p(x)U_{i}\left(x+a\frac{\hat{i}}{2},0\right)p(x+a\hat{i})^{\dagger}$%
  \State $U_{t}\left(x,0\right) \gets p(x)U_{t}\left(x,0\right)q(x)^{\dagger}$%
  \For {$t=2 \rightarrow L_t-a$}
    \State $n = 0, r(x) \gets I_{N_c\times N_c}$%
    \While {$\theta^{(n)}(t) > \text{Tolerance} $}
    \State $r(x) \gets  e^{-i\alpha \: \tilde{F} \: \frac{p^2_{\text{Max}}}{V_{N_d-1}p^2} \: F \: a\Delta_i ag_0 A_i^{(n)}(x,t) }r(x)$%
    \State $n = n+1$
    \EndWhile
    \State $U_{i}\left(x+a\frac{\hat{i}}{2},t-a\right) \gets q(x)U_{i}\left(x+a\frac{\hat{i}}{2},t-a\right)q(x+a\hat{i})^{\dagger}$%
    \State $U_{t}\left(x,t-a\right) \gets q(x)U_{t}\left(x,t-a\right)r(x)^{\dagger}$%
    \State $q(x) \gets r(x)$
  \EndFor
  \State $U_{i}\left(x+a\frac{\hat{i}}{2},L_t-a\right) \gets q(x)U_{i}\left(x+a\frac{\hat{i}}{2},L_t-a\right)q(x+a\hat{i})^{\dagger}$%
  \State $U_{t}\left(x,L_t-a\right) \gets q(x)U_{t}\left(x,L_t-a\right)p(x)^{\dagger}$%
  \end{algorithmic}
\end{algorithm}

Occasionally, the algorithm on a time-slice converges slowly. This is due to the initial topography of the fields, and is remedied by performing a random gauge transformation on the initial link matrices if $n$ is larger than some predetermined number and $\theta^{(n)}(t)$ is not close to the desired accuracy. 

Storing our gauge transformation matrices as an accumulated product can lead to small round off errors, which we compensate for by reunitarisation before applying the gauge transformation (Eq.\ref{eq:gtrans_method}) to the links.

The extension of the time-slice by time-slice FASD routine outlined in Alg.\ref{Alg:CFASD} to a FACG routine requires the replacement of each while loop that sets the spatial gauge transformation matrices $p,q$ and $r$ with a spatial-polarisation variant of Alg.\ref{alg:CG_FALAN}. As in the FASD case, we compute the accumulated product of the gauge transformation matrices instead of overwriting the links at each step for our Coulomb gauge FACG method.

\section{Results}\label{sec:results}

\begin{figure}[h!]
\centering
  {
  \includegraphics[scale=0.33]{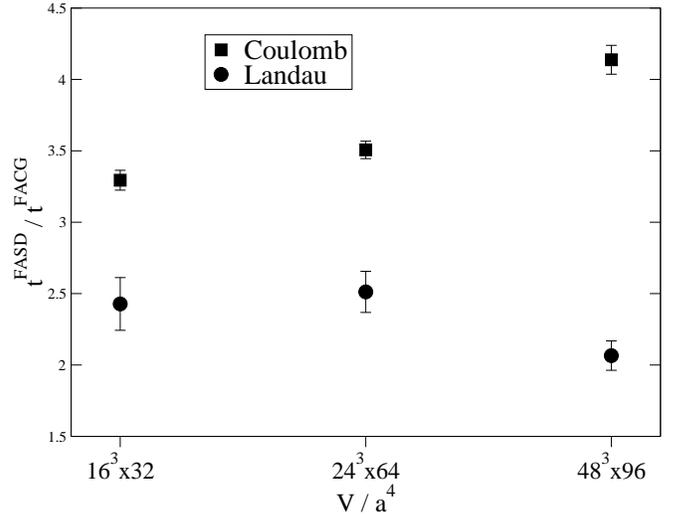}
  }
\caption[]{The improvement factor $\frac{t^{\text{FASD}}}{t^{\text{FACG}}}$ defined by the ratio of the time taken to achieve a gauge fixing accuracy of $\Theta^{(n)} < 10^{-14}$ for both the Landau and Coulomb FASD to FACG routines.}%
\label{fig:speedups}
\end{figure}

Fig.\ref{fig:speedups} illustrates a measure of the effective speed up for our conjugate gradient approach by showing the average time taken to achieve some common accuracy using the FASD algorithm at near-optimal fixed step $\alpha$, divided by the time taken to achieve the same accuracy for the FACG algorithm for both Landau and Coulomb gauge. The results are generated from the same ($\text{SU}(3),N_d=4$) gauge configuration, randomly gauge transformed 25 times, for 3 different Iwasaki gauge action DWF ensembles with the same lattice spacing of $a^{-1}\approx 1.75$ GeV \cite{Arthur:2012opa} and different physical volumes.

The time taken to achieve a fixed accuracy is a more representative measurement of the real-world application of this procedure than say the time taken for the algorithm to perform a set number of iterations. This is because the number of iterations required to achieve fixed accuracy grows with volume (a sure sign that the Fourier acceleration is not removing all critical slowing down). Although one iteration should either scale like $O(V\log(V))$ or $O(V)$ depending on whether the cost in performing the FFT is dominant or not, the total time taken to achieve a fixed accuracy is measured to grow like $V^n$ where $1<n<2$.

\begin{table}[h]
\begin{tabular}{|c|c|c|c|c|}
\hline
& \multicolumn{2}{|c|}{Landau} & \multicolumn{2}{|c|}{Coulomb} \\
\hline
Volume & FASD (s) & FACG (s) & FASD (s) & FACG (s)\\
\hline
$16^3\times32$ & 25(1) & 10.5(5) & 27.4(6) & 8.3(1) \\
$24^3\times64$ & 306(18) & 122(7) & 194(3) & 55.4(4) \\
$48^3\times96$ & 12995(857) & 6292(404) & 10882(180) & 2629(39) \\
\hline
\end{tabular}
\caption{The time in seconds taken to achieve an accuracy of $\Theta^{(n)} < 10^{-14}$ for 25 randomly transformed copies of the same, well-thermalised configuration. Each configuration has the same lattice spacing. This measurement was performed using four 8-core AMD Opteron 6128 processors. All operations were performed in double precision.}\label{tab:data}
\end{table}

From Fig.\ref{fig:speedups} we see that the algorithm gives roughly a $\approx3.5\times$ speed up over our FASD implementation for Coulomb gauge and better performance going to larger volumes. For Landau gauge we achieve a more modest $\approx2.5\times$, with a slight drop in performance as we go to large volumes. Tab.\ref{tab:data} gives the raw data for Fig.\ref{fig:speedups}, and we see that even for large lattices the problem is particularly tractable even on small clusters of CPUs.

\section{Conclusions}\label{sec:conclusions}

Fixing the gauge is often necessary for lattice analyses. It should not, however, be a computational bottleneck as fast algorithms with good scaling properties exist. Choosing the best algorithm should take precedence over the use of expensive hardware and aggressive optimisations when implementing a gauge fixing procedure.

We have introduced a new method for fixing to Landau and Coulomb gauge that is shown to be faster by a factor of more than 2 to 4 times respectively over the commonly used Fourier accelerated steepest descent method, albeit for a very moderate increase in required computer memory. We conclude that the upgrade from a FASD to a FACG routine is straightforward and worthwhile.

Implementations of the FASD and FACG algorithm for both Landau and Coulomb gauge are available as part of the package GLU, which is an open source shared-memory parallel gluonic observable library and is available for download at 
\href{https://github.com/RJhudspith/GLU}{https://github.com/RJHudspith/GLU}.

\section{Acknowledgements}

The results produced in this work were generated on Columbia University's cluster ``CUTH''. RJH would like to thank M.Wurtz, N. Garron and R. Lewis for providing useful comments about this manuscript. RJH is supported by the Natural Sciences and Engineering Research Council of Canada.

\bibliographystyle{JHEP}
\bibliography{Landau_CG_pCoul}

\providecommand{\href}[2]{#2}\begingroup\raggedright\begin{thebibliography}{10}

\bibitem{zbMATH03075195}
M.~R. {Hestenes} and E.~{Stiefel}, {\it {Methods of conjugate gradients for
  solving linear systems.}},  {\em {J. Res. Natl. Bur. Stand.}} {\bf 49} (1952)
  409--436.

\bibitem{Fletcher01011964}
R.~Fletcher and C.~M. Reeves, {\it Function minimization by conjugate
  gradients},  {\em The Computer Journal} {\bf 7} (1964), no.~2 149--154.

\bibitem{Polyak196994}
B.~Polyak, {\it {The conjugate gradient method in extremal problems}},  {\em
  \{USSR\} Computational Mathematics and Mathematical Physics} {\bf 9} (1969),
  no.~4 94–112.

\bibitem{Press:1992:NRC:148286}
W.~H. Press, S.~A. Teukolsky, W.~T. Vetterling, and B.~P. Flannery, {\em
  Numerical recipes in C (2nd ed.): the art of scientific computing}.
\newblock Cambridge University Press, New York, NY, USA, 1992.

\bibitem{Aoki:2010pe}
Y.~Aoki, R.~Arthur, T.~Blum, P.~Boyle, D.~Brommel, et~al., {\it {Continuum
  Limit of $B_K$ from 2+1 Flavor Domain Wall QCD}},  {\em Phys.Rev.} {\bf D84}
  (2011) 014503, [\href{http://xxx.lanl.gov/abs/1012.4178}{{\tt
  arXiv:1012.4178}}].

\bibitem{Boyle:2012qb}
{\bf RBC Collaboration, UKQCD Collaboration} Collaboration, P.~Boyle,
  N.~Garron, and R.~Hudspith, {\it {Neutral kaon mixing beyond the standard
  model with $n_f = 2+1$ chiral fermions}},  {\em Phys.Rev.} {\bf D86} (2012)
  054028, [\href{http://xxx.lanl.gov/abs/1206.5737}{{\tt arXiv:1206.5737}}].

\bibitem{Blossier:2013ioa}
{\bf ETM Collaboration} Collaboration, B.~Blossier et~al., {\it {High
  statistics determination of the strong coupling constant in Taylor scheme and
  its OPE Wilson coefficient from lattice QCD with a dynamical charm}},  {\em
  Phys.Rev.} {\bf D89} (2014) 014507,
  [\href{http://xxx.lanl.gov/abs/1310.3763}{{\tt arXiv:1310.3763}}].

\bibitem{Kaczmarek:2002mc}
O.~Kaczmarek, F.~Karsch, P.~Petreczky, and F.~Zantow, {\it {Heavy quark
  anti-quark free energy and the renormalized Polyakov loop}},  {\em
  Phys.Lett.} {\bf B543} (2002) 41--47,
  [\href{http://xxx.lanl.gov/abs/hep-lat/0207002}{{\tt hep-lat/0207002}}].

\bibitem{Mandula1987127}
J.~Mandula and M.~Ogilvie, {\it The gluon is massive: A lattice calculation of
  the gluon propagator in the landau gauge},  {\em Physics Letters B} {\bf 185}
  (1987), no.~1–2 127 -- 132.

\bibitem{Durr:2007cy}
S.~Durr, {\it {Logarithmic link smearing for full QCD}},  {\em
  Comput.Phys.Commun.} {\bf 180} (2009) 1338--1357,
  [\href{http://xxx.lanl.gov/abs/0709.4110}{{\tt arXiv:0709.4110}}].

\bibitem{Ilgenfritz:2010gu}
E.-M. Ilgenfritz, C.~Menz, M.~Muller-Preussker, A.~Schiller, and A.~Sternbeck,
  {\it {SU(3) Landau gauge gluon and ghost propagators using the logarithmic
  lattice gluon field definition}},  {\em Phys.Rev.} {\bf D83} (2011) 054506,
  [\href{http://xxx.lanl.gov/abs/1010.5120}{{\tt arXiv:1010.5120}}].

\bibitem{PhysRevD.37.1581}
C.~T.~H. Davies, G.~G. Batrouni, G.~R. Katz, A.~S. Kronfeld, G.~P. Lepage,
  K.~G. Wilson, P.~Rossi, and B.~Svetitsky, {\it Fourier acceleration in
  lattice gauge theories. i. landau gauge fixing},  {\em Phys. Rev. D} {\bf 37}
  (Mar, 1988) 1581--1588.

\bibitem{PhysRevD.69.054501}
C.~Morningstar and M.~Peardon, {\it Analytic smearing of $su(3)$ link variables
  in lattice qcd},  {\em Phys. Rev. D} {\bf 69} (Mar, 2004) 054501.

\bibitem{FFTW05}
M.~Frigo and S.~G. Johnson, {\it The design and implementation of {FFTW3}},
  {\em Proceedings of the IEEE} {\bf 93} (2005), no.~2 216--231. Special issue
  on ``Program Generation, Optimization, and Platform Adaptation''.

\bibitem{Cucchieri:1998ew}
A.~Cucchieri and T.~Mendes, {\it {A Multigrid implementation of the Fourier
  acceleration method for Landau gauge fixing}},  {\em Phys.Rev.} {\bf D57}
  (1998) 3822--3826, [\href{http://xxx.lanl.gov/abs/hep-lat/9711047}{{\tt
  hep-lat/9711047}}].

\bibitem{Nakagawa:2009zf}
Y.~Nakagawa, A.~Voigt, E.-M. Ilgenfritz, M.~Muller-Preussker, A.~Nakamura,
  et~al., {\it {Coulomb-gauge ghost and gluon propagators in SU(3) lattice
  Yang-Mills theory}},  {\em Phys.Rev.} {\bf D79} (2009) 114504,
  [\href{http://xxx.lanl.gov/abs/0902.4321}{{\tt arXiv:0902.4321}}].

\bibitem{Arthur:2012opa}
{\bf RBC Collaboration, UKQCD Collaboration} Collaboration, R.~Arthur et~al.,
  {\it {Domain Wall QCD with Near-Physical Pions}},
  \href{http://xxx.lanl.gov/abs/1208.4412}{{\tt arXiv:1208.4412}}.

\end{thebibliography}\endgroup

\end{document}